\documentclass[preprint,12pt]{elsarticle}

\usepackage{hyperref}
\usepackage{color, soul}

\usepackage{amssymb}
\usepackage{amsmath}

\newcounter{bla}

\journal{Computer Physics Communications}

\begin{document}

\begin{frontmatter}



\title{{\bf O}TSLM {\bf T}oolbox for {\bf S}tructured {\bf L}ight {\bf M}ethods}


\author{Isaac C D Lenton$^*$}
\author{Alexander B Stilgoe}
\author{Timo A Nieminen}
\author{Halina Rubinsztein-Dunlop}

\cortext[author] {Corresponding author. \textit{E-mail address:} i.lenton@uq.edu.au}
\address{School of Mathematics and Physics,
    The University of Queensland, St Lucia, QLD 4072, Australia}

\begin{abstract}
We present a new Matlab toolbox for generating phase and amplitude patterns for
digital micro-mirror device (DMD) and liquid crystal (LC) based spatial light
modulators (SLMs).
This toolbox consists of a collection of algorithms commonly used for
generating patterns for these devices with a focus on optical tweezers beam
shaping applications.
In addition to the algorithms provided, we have put together a range of user
interfaces for simplifying the use of these patterns.
The toolbox currently has functionality to generate patterns which can be
saved as a image or displayed on a device/screen using the supplied interface.
We have only implemented interfaces for the devices our group currently uses
but we believe that extending the code we provide to other devices should be
fairly straightforward.
The range of algorithms included in the toolbox is not exhaustive.
However, by
making the toolbox open sources and available on GitHub we hope that other
researchers working with these devices will contribute their patterns/algorithms
to the toolbox.
\end{abstract}



\end{frontmatter}

\textbf{Pre-print of:}
\begin{quote}
    I. Lenton, et al., 
    OTSLM toolbox for Structured Light Methods,
    Computer Physics Communications,
    2020, 107199,
    \url{https://doi.org/10.1016/j.cpc.2020.107199}.
    
\copyright~2019. This manuscript version is made available under the CC-BY-NC-ND 4.0\\
license \url{http://creativecommons.org/licenses/by-nc-nd/4.0/}.
\end{quote}



{\bf PROGRAM SUMMARY}

\begin{small}
\noindent
{\em Program Title: OTSLM Toolbox for Structured Light Methods} \\
{\em Licensing provisions: GPLv3}                               \\
{\em Programming language: Matlab}                              \\
{\em Nature of problem:} \\ There are many algorithms for generating computer
  controlled holograms however the code and descriptions for these algorithms
  is often provided as supplementary material to research publications which
  use these methods, and in some cases only the description is provided without code
  to reproduce the pattern.
  Furthermore, implementation of these algorithms can be a time consuming task.
  Even for simple patterns with simple analytical expressions for the far-field
  phase or amplitude, such as the Laguerre-Gaussian and Hermite-Gaussian beams,
  time is often wasted by researchers having to re-implement and test these patterns.
  Existing libraries for generation of SLM patterns focus on specific tasks,
  methods of pattern generation, or target specific hardware.
  These libraries are not general purpose or easily modifiable for
  general use by researchers working with computer controlled holograms in
  fields such as beam shaping and optical tweezers.
  \\  
{\em Solution method:} \\ We have assembled a toolbox containing many
  methods commonly used with these devices.  The number of methods available
  makes assembling a complete toolbox impossible, instead we have
  focused on putting together a general collection of methods, in the form of
  an open source toolbox, with the hope that other researchers
  will contribute patterns/algorithms they use.
  The toolbox currently includes a range of simple and iterative methods
  for beam shaping and steering as well as some of the methods our group currently
  uses for SLM control, calibration, imaging and optical tweezers beam shaping.
  To make the tools we have developed easy to use, we have tried to maintain
  a consistent well documented interface to each of the functions and provided
  graphical user interfaces to many of the simpler functions enabling code-free
  pattern generation.
  \\  
{\em Additional comments:}\\
  Some of the features require the Optical Tweezers Toolbox [1], in particular, the
  non-paraxial beam visualisation and optimisation routines.
  Certain functions require components from other Matlab packages such as the
  image acquisition toolbox, these can be licensed and installed from Mathworks.
  There is also interest from the authors in providing a Python version that
  requires no proprietary components depending on interest from the community.
  A public repository for the toolbox is available on GitHub [2].\\

\end{small}

\section{Introduction}
\label{sec:introduction}

Beam shaping using computer controlled spatial light modulators is useful in many
fields with applications in imaging, optical tweezers, and optical fabrication.
In optical tweezers, beam shaping is used to generate the beams used to optically
confine the particles under study \cite{Grier2006Feb}, construct elaborate imaging
systems \cite{Maurer2011Jan, Stilgoe2018Mar}, and for fabricating probes and
structures that can be optically trapped.
A modern optical tweezers system typically consists of a laser, a system for shaping
the laser beam, a high numerical aperture objective, the sample, a condenser and the
imaging/detection system.
Such a system enables numerous studies to be conducted
in highly versatile fields including biology,
physics and engineering \cite{Padgett2011May, Maurer2011Jan}.
Optical tweezers are also being used in quantum opto-mechanics in studies
concerned with borderline between classical and quantum physics.
In all of these applications beam shaping is essential in enabling desired studies.  
Despite the extensive use of computer controlled holograms in the field of optical
tweezers, there is limited availability of code for generating the types of patterns
typically used for these applications.
To address this issue, we have put together a toolbox of functions useful for optical
tweezers researchers, however we believe that the toolbox will be more broadly useful
to other fields involving beam shaping and structured light with theses devices.


Existing libraries for generation of SLM patterns focus on specific tasks or methods of
pattern generation.
Tools for optical tweezers include red-tweezers \cite{Bowman2014Jan}, written in LabView for
generating patterns using computer graphics cards; and HoloTrap \cite{Pleguezuelos2007Jun},
written in Java for real time generation of binary mask holograms.
There are also tools for specific hardware, such as HOLOEYE SLM Pattern Generator \cite{Holoeye2018Jul},
and tools with other intended applications such as llspy-slm for
light sheet microscopy \cite{llspy}.
Some groups make available the code they use to demonstrate a particular
algorithm or result for a paper, such as \cite{slm-cg}, however these
published codes are sometimes difficult to find/access, are rarely user friendly
and often need additional modification for integration into an optical tweezers system.


The toolbox we have developed \cite{OtslmGithub}, does not focus on a particular
device and includes patterns of interest to the broad optical trapping community
and hopefully the wider beam shaping community.
At this stage we include the patterns that we have used in our recent and ongoing
experiments. 
To make the toolbox easy to use, we have provided both a Matlab programmers interface
and a graphical user interface allowing the tools to be used without the need to write code.
We have tried to ensure that the toolbox follows a consistent documentation format and
tried to keep the interfaces similar between similar functions to make it easier to move
between different components in the toolbox.
By making the toolbox open source and available on GitHub, it is hoped that the
wider computer controlled hologram community will contribute their codes/methods for
generating patterns they use and make them more accessible.
It is our hope that this will lead to a curated toolbox of functions for beam
shaping with a consistent user interface and documentation that will be
beneficial to the community.

\section{Background}

\subsection{Device overview and operation}

There are a variety of systems used for beam shaping, but among the most common
are amplitude and phase spatial light modulators, such as the digital micro-mirror device
(binary amplitude SLM, or DMD) or liquid crystal (phase SLM) \cite{woerdemann2013, dholakia2011}\footnote{It is important to note that, in the literature, the term SLM may be used ambiguously to refer to either phase-only devices or any device capable of modulating the beam wave-front. In this article we use SLM to refer to any device capable of modulating the beam phase and/or amplitude.}
These highly configurable devices can be controlled using a variety of
different interfaces but a common method is to connect them as an additional
screen on the lab computer. The image displayed on the screen corresponds
to the amplitude or phase pattern displayed on the device.
There are numerous guides for setting up one of these devices in an optical tweezers
system \cite{Jones2015Dec, Pesce2015May}, and many of these guides
describe generating SLM patterns for beam shaping but they do not provide
easy to use codes. In some cases \cite{Pesce2015May}, tasks such as
simultaneous phase/amplitude modulation are not described.

\begin{figure}[ht]
  \centering
  \includegraphics[width=0.6\textwidth]{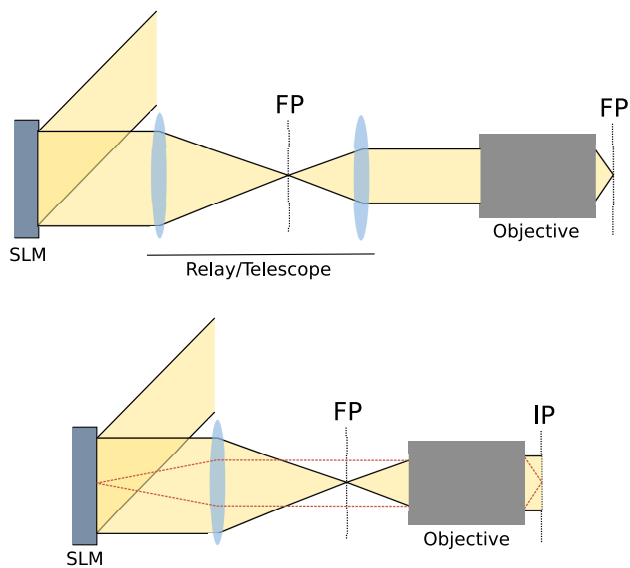}
  \caption{Two configurations for phase/amplitude SLM used in a microscope system.
  The device is placed so the objective plane corresponds to a
  Fourier plane (FP) of the device (top) and an imaging plane (IP)
  of the device (bottom).
  The relay/telescope is optional and
  provides access to the Fourier plane which is useful for removing
  non-diffracted light or spatially filtering the beam.}
  \label{fig:setup}
\end{figure}

For creating structured optical tweezers beams, the incident beam either reflects or
passes through the device before being focused by a microscope objective.
The position of the device with respect to the microscope objective changes how
the device controls the beam. Two commonly used configurations are shown in
figure \ref{fig:setup}.
Phase based devices are typically imaged onto the back focal plane of the
objective to allow spatial control of the generated pattern at the focal plane.
When the device is imaged onto the back focal plane, the focal
plane corresponds to a conjugate plane of the SLM, i.e. the image at the objective
focal plane corresponds to the spatial frequency spectrum of the image on the device.
Amplitude based devices can be imaged onto the back plane, producing a
Fourier image at the focal plane \cite{lerner2012, Porfiriev2014Jul},
or imaged directly to the focal plane \cite{gauthier2016}.
When imaged directly, the beam generated by amplitude based devices is somewhat
intuitive, since the pattern on the SLM corresponds to the pattern at the focal
plane with resolution limited by aberrations and the transfer function of the system.
Intermediate planes between the imaging and conjugate planes can be thought of as a mixture
of the image and conjugate planes.
In addition to the two configurations described here, other systems can
involve placing the SLM in a Fresnel plane or other plane \cite{Padgett2011}
or omitting the lens entirely \cite{Fixler2011Oct}.
It is also possible to use multiple devices or multiple passes of the same device
to achieve greater control over both amplitude and phase \cite{Jesacher2008Mar}.

Determining the appropriate pattern to place on the device to generate the
desired beam depends on a number of factors including the device position
and incident illumination.
If the device is positioned in the conjugate plane and uniformly illuminated,
the required device pattern is simply the far-field complex amplitude of the desired
beam.
For simple beam shapes, such as beams translated to different positions,
the required patterns can be easily determined and correspond to patterns for
diffraction gratings and lenses.
Once the complex amplitude function for the desired beam is determined,
the phase or amplitude pattern can be calculated and converted to a device specific
image with values corresponding to the phase/amplitude range of the device.
For non-uniform incident illumination, pattern generation becomes more
complex: the beam phase and amplitude distribution and the phase uniformity
of the device is needed, which must then be encoded in the generated pattern.
More complex patterns, where the far-field phase or intensity is not intuitive,
can be generated using iterative algorithms such as
the popular Gerchberg--Saxton algorithm \cite{Gerchberg1971Nov}.

When using one of these devices there are several additional considerations
arising from the physical properties of the device that may impact the generated patterns.
For example, for phase-only SLMs, such as liquid crystal type devices,
the orientation of the liquid crystals is controlled by applying a voltage
to align the crystals in a particular direction.
The change in orientation of the liquid crystals results in a
change in phase and if the crystals
are birefringent, a change in polarisation of light passing through the
corresponding region of the device will occur (making the device not purely phase-only).
Part of the setup for one of these devices often involves calibrating the
device voltage to corresponding phase/polarisation values.
The physical process involved in changing the phase/polarisation of the light
may not be perfectly efficient, this leads to effects such as pixel blurring
and non-diffracted light in the resulting beam.

\subsection{Simple patterns}

For optical tweezers experiments, there are a range of simple patterns that are
often used for basic beam steering and control of trapped particles.
The most common optical tweezers beam
is a Gaussian beam, which can be
described in the far-field of the focus by a scalar amplitude function
with constant phase and a Gaussian profile,
\begin{equation}
    u_{gaussian}(x, y) \propto \exp\left(-\frac{x^2 + y^2}{2\sigma^2}\right)
\end{equation}
with only a single parameter, $\sigma$, the width of the beam.

\begin{figure}
\centering
\makebox[15px][r]{a) }\includegraphics[width=2cm]{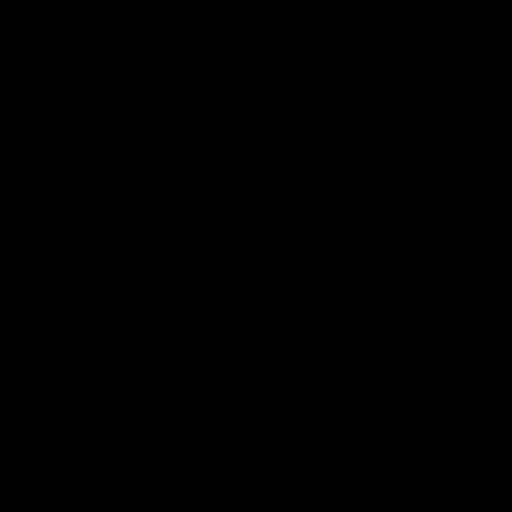}%
\includegraphics[width=2cm]{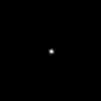}
\makebox[15px][r]{e) }\includegraphics[width=2cm]{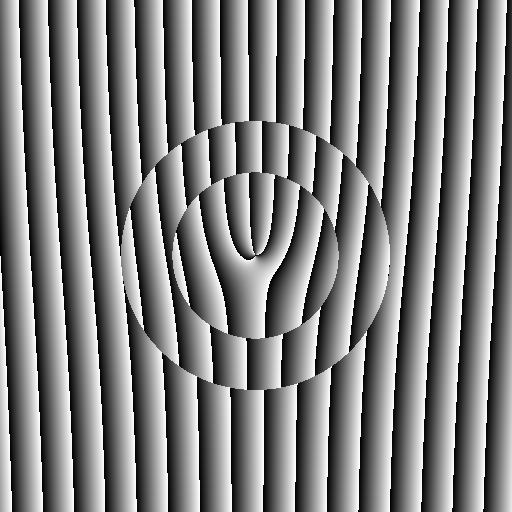}%
\includegraphics[width=2cm]{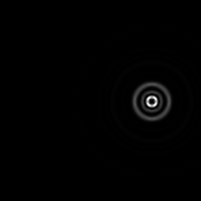}
\makebox[15px][r]{b) }\includegraphics[width=2cm]{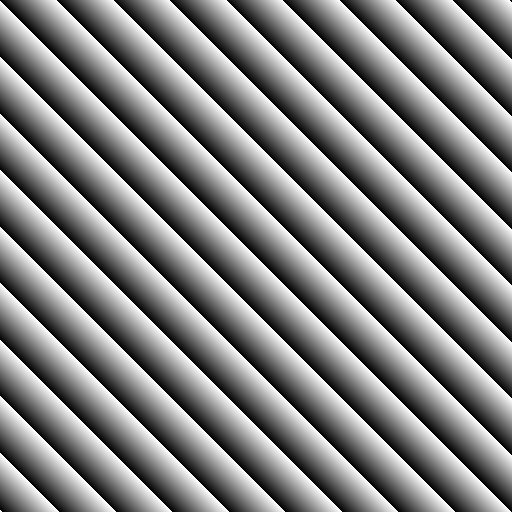}%
\includegraphics[width=2cm]{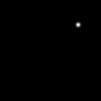}
\makebox[15px][r]{f) }\includegraphics[width=2cm]{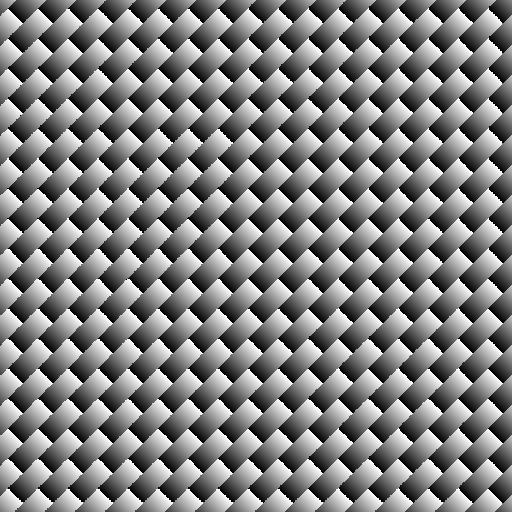}%
\includegraphics[width=2cm]{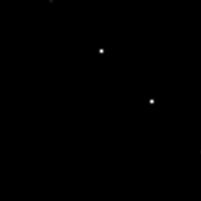}
\makebox[15px][r]{c) }\includegraphics[width=2cm]{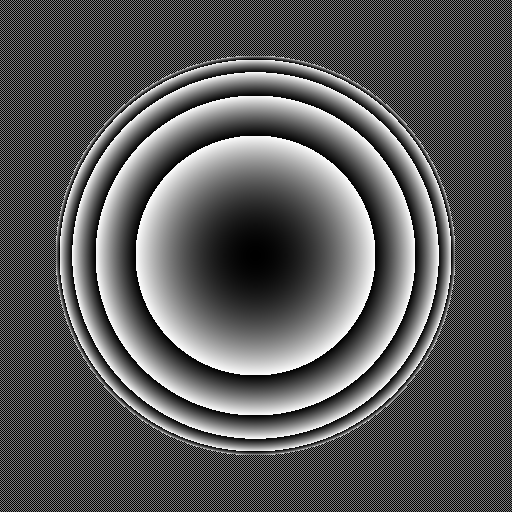}%
\includegraphics[width=2cm]{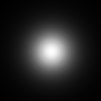}
\makebox[15px][r]{g) }\includegraphics[width=2cm]{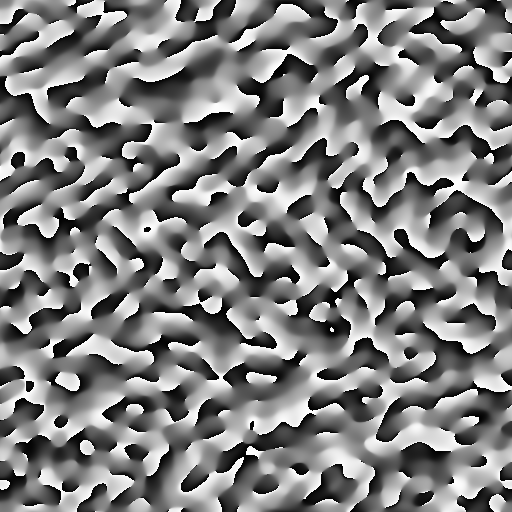}%
\includegraphics[width=2cm]{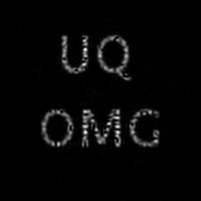}
\makebox[15px][r]{d) }\includegraphics[width=2cm]{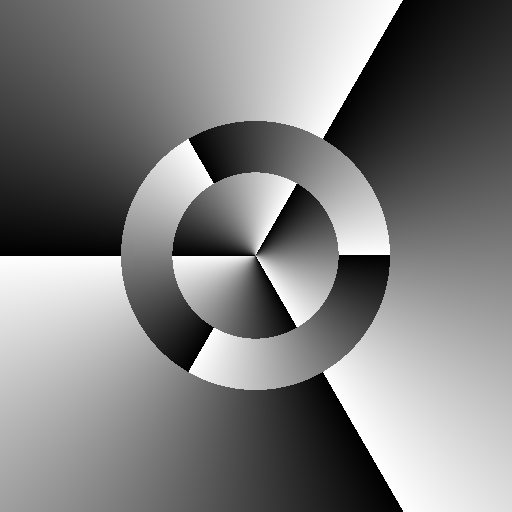}%
\includegraphics[width=2cm]{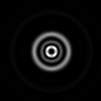}
\makebox[15px][r]{h) }\includegraphics[width=2cm]{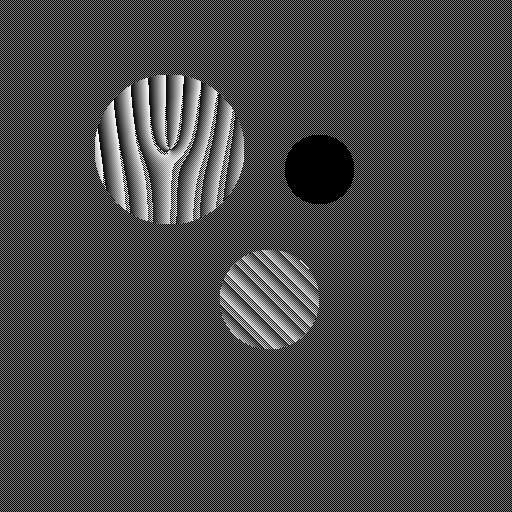}%
\includegraphics[width=2cm]{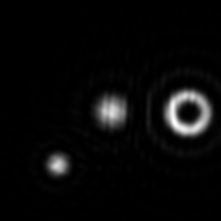}
\caption{Phase patterns generated using the toolbox (left) and visualisations
  of the SLM far-field when the pattern is illuminated by a Gaussian beam (right):
  (a) zero phase, (b) blazed (linear) diffraction grating, (c) spherical lens,
  (d) LG mode pattern, (e) LG mode displaced with linear grating,
  (f) two superimposed linear gratings,
  (g) results of Gurchberg-Saxton algorithm after a few iterations,
  (h) sampling of different SLM regions.
  (c) and (h) use a checkerboard over regions outside the
  pattern to scatter incident light to higher angles.}
\label{fig:beams}
\end{figure}

We can focus or deflect a beam by applying a lens or linear phase grating to the beam.
To first order approximation the phase profiles for a lens and grating are given by
\begin{eqnarray}
    \phi_{lens}(x, y) &=& \frac{x^2 + y^2}{2R} \\
    \phi_{grating}(x, y) &=& D(x\sin(\theta) + y\cos(\theta))
\end{eqnarray}
where $R$ is the radius of curvature for the lens, $D$ is the distance
displaced and $\theta$ is the direction of the displacement.
By applying these phase shifts to our Gaussian beam, we are able to move the focus
of the beam in the axial direction
$\phi_{lens}$ and in the transverse direction $\phi_{grating}$.
These formulas can be applied to a beam by multiplying the beam by the corresponding
complex field amplitude
\begin{equation}
    u_{translated}(x, y) = \exp\left(ik (\phi_{lens}(x,y) + \phi_{grating}(x,y))\right) u_{original}(x,y)
\end{equation}
where $i$ is the imaginary unit and $k$ is the optical wavenumber.
Examples of these diffraction gratings are shown in figure \ref{fig:beams}(a-c).

Orbital angular momentum can be added to a beam by adding a azimuthal phase
gradient to the pattern.
The Laguerre-Gaussian modes are Gaussian beam modes which include a
azimuthal phase component, this additional phase is described by
\begin{eqnarray}
    \phi_{LG_{0l}} = l\arctan(y/x).
\end{eqnarray}
Figure \ref{fig:beams}(d-e) show LG modes with both azimuthal phase and radial phase.
More complex patterns can be generated for Hermite-Gauss, Ince-Gauss and
other types of beams.

\subsection{Combining multiple patterns}

It may be desirable to combine a set of patterns together to create multiple
beams.
There are multiple methods of achieving this including the random mask
method and superposition of prisms and lenses.
The random mask method involves randomly choosing pixels from each pattern
and combining them into a single image.
To combine a set of patterns $\phi_i$ using the random mask method,
the pixels in the combined pattern are given by
\begin{equation}
  \phi_{combined}^j(x, y) = \phi_i^j(x, y)
\end{equation}
where $j$ refers to the individual pixels that make up the pattern and
$i$ is chosen randomly for each pixel.
By changing the frequency at which a particular pattern is chosen,
the amplitude of the individual patterns can be controlled.
The efficiency for this method for uniform beam amplitudes scales inversely
with the number of traps, providing lower efficiency compared to other methods.

The superposition of prisms and lenses algorithm involves superimposing the
complex amplitude patterns of each pattern and calculating the phase angle
\begin{equation}
    \phi_{combined}(x, y) = \arg\left(\sum_{i} w_i e^{i\phi_i(x, y) + \theta_i}\right).
\end{equation}
The additional weighting terms, $w_i$, are used to control the amplitude of
each trap, while the phase offset terms, $\theta_i$ can be used to introduce
a random phase offset which may improve the performance of
the method \cite{DiLeonardo2007Feb}.
For further reading see \cite{DiLeonardo2011Jan}.
An example of superposition is shown in figure \ref{fig:beams}(f).

\subsection{Simulating tightly focused beams}

\begin{figure}
\centering
\includegraphics[width=\textwidth]{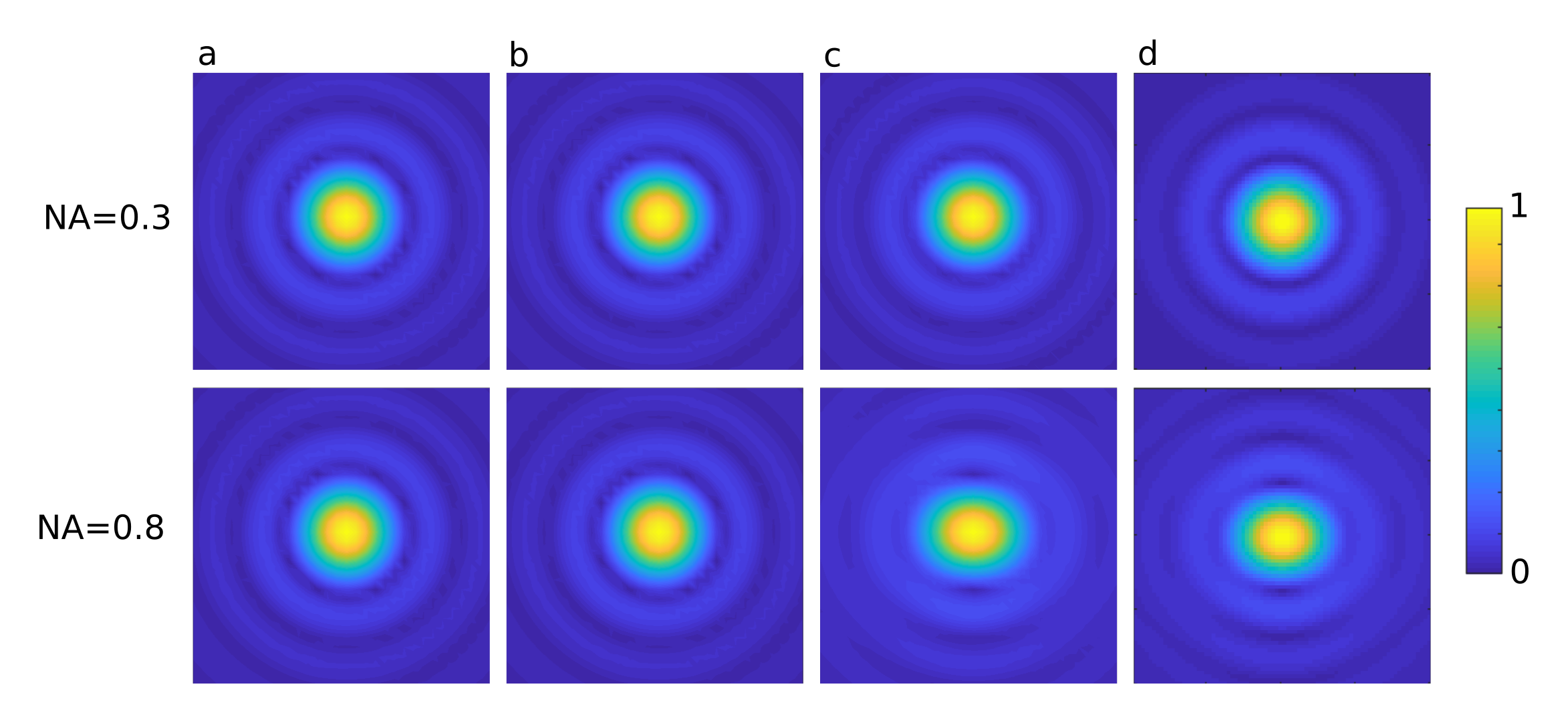}
\caption{Comparison of different simulation methods when simulating
    a weakly focused (NA=0.3) and tightly focused (NA=0.8) linearly
    polarised Gaussian beam.
    Simulation show the irradiance of the focused beam for
    (a) scalar 2-D FFT,
    (b) scalar 3-D FFT,
    (c) vector 2-D FFT implementation of Debye integral,
    and (d) point-matching to vector spherical wave functions using
    the optical tweezers toolbox.}
\label{fig:vis_methods}
\end{figure}

Optical tweezers often use tightly focused beams created using
high numerical aperture (NA) objectives (NA $\sim$ 0.7--1).
These beams are often not solutions to the scalar Helmholtz equation.
Full modelling often requires higher order approximations or solutions
to the vector Helmholtz equation \cite{Peatross2017Jun, Vaveliuk2007Oct}.
The simulations in figure~\ref{fig:beams} assumed the beams to be
paraxial (scalar)
and do not account for changes to the beam polarisation due to focusing.
For low numerical aperture simulations, this can produce reasonably accurate
results.
However, for beams used in optical tweezers it can be necessary to include
polarisation effects.
Figure~\ref{fig:vis_methods} shows a comparison of four different beam
simulation methods currently implemented in OTSLM
\citep{Delen1998Apr, Shabtay2003Oct, Leutenegger2006Nov, ottv1}.
For weakly focused beams (NA=0.3) there is very little visible difference
between the different methods.
The scalar fast Fourier transform (FFT) methods,
figure~\ref{fig:vis_methods}(a--b), do not include the
polarisation of the incident beam and produces a spot which can differ
significantly from the experimentally realised beams.
In order to model polarisation effects in highly focused beams, one simple
method is to repeat the FFT method for each polarisation component.
Further adaptations can be added to account for spherical aberration due
to the lenses finite dimensions and the projection of the paraxial
polarisation components onto the hemisphere of the lens \citep{Leutenegger2006Nov}.

\subsection{Simultaneous amplitude \& phase control}

In order to generate arbitrary beams both the amplitude and phase of
the wave-front need to be controlled.
Many liquid crystal type SLMs can be configured so that they control 
the phase and amplitude of the reflected light, however the phase and
amplitude control is rarely independent \cite{deBougrenetdelaTocnaye1997Mar}.
Multiple devices can be combined, a device that only
controls phase can be combined with a device that only controls amplitude:
such as combing a liquid crystal type SLM with a DMD type SLM or two
liquid crystal devices setup to modulate the phase and amplitude respectively \cite{Zhu2014Dec}.
If spatial resolution isn't important, the same device can be used in
a double pass configuration with the device being split into two halves
one for controlling amplitude and the other for phase.
It is also possible to use a spatial filter to combine adjacent pixels
into a super-pixel that covers the complete amplitude/phase
space \cite{vanPutten2008Apr, Goorden2014Jul}.
More complicated systems can even achieve phase, amplitude and polarisation
control simultaneously with only a single DMD, half waveplates and beam
displacers \cite{Mitchell2016Dec}.

Without additional hardware or additional devices it is still possible
to modify the SLM pattern to take advantage of the available
amplitude/phase space.
For phase only devices, one approach is to
mix the desired phase pattern with a second phase pattern
to give the phase pattern with both amplitude and phase information
\begin{equation}
    \phi_{A+P} = f(A)\phi_P + (1 - f(A))\phi_{other}
\end{equation}
where $f(A)$ is a function between 0 and 1 which converts the
amplitude pattern into a mixing ratio and $\phi_{other}$ is the
other phase pattern to mix in.
There are multiple choices for the function $f$, the simplest situation
is $f(A) = A$ and to discard light into the zero-th order so the above
equation becomes
\begin{equation}
    \phi_{A+P} = A\phi_P.
\end{equation}
For a comparison of different methods involving mixing multiple patterns
to encode the amplitude on a phase-only device, see \cite{Clark2016Mar}.
A similar procedure can be done for amplitude-only device \cite{lerner2012}.

\subsection{Iterative algorithms}

For generating more complex patterns there are a range of
iterative algorithms that attempt to generate the incident phase
or amplitude distribution for a desired target distribution
for both phase only \cite{courtial2006, DiLeonardo2007Feb, Poland2014Apr}
and amplitude only devices \cite{stuart2014}.
One of the most popular and simple methods is the 2-D
Gerchberg--Saxton (GS) algorithm \cite{Gerchberg1971Nov}, shown bellow:
\begin{enumerate}
    \item Generate initial guess at the SLM phase pattern: $P$
    \item Calculate output for phase pattern: $\text{Proj}(P) \rightarrow O$ \label{item:gs-start}
    \item Multiply output phase by target amplitude: $|T|\frac{O}{|O|} \rightarrow Q$
    \item Calculate the complex amplitude required to generate $Q$: $\text{Inv}(Q) \rightarrow I$
    \item Calculate new guess from the phase of $I$: $\text{Angle}(I) \rightarrow P$
    \item Goto step \ref{item:gs-start} until converged
\end{enumerate}
Proj is an operation to calculate the output field and Inv is
the corresponding inverse operation, in the original GS these are the Fourier
and inverse Fourier transforms.
There are a range of modifications to this algorithm which achieve better
uniformity or compatibility with amplitude only devices.
An example of the GS algorithm is shown in figure \ref{fig:beams}(g).
The 2-D Gerchberg--Saxton algorithm can be used to generate 3-D holograms
by constraining the output to multiple planes in 3-D space or using a
modified 3-D algorithm which allows optimisation with respect to a target
volume \cite{Chen2013Jan, Whyte2005May}.

Other algorithms include direct search and simulated annealing which involves
randomly changing pixel values to try and optimise the
pattern \cite{Yoshikawa1994Feb, spalding2008holographic};
and gradient descent algorithms which attempt to directly optimise some
fitness function \cite{Bowman2017May}.

It is also possible to optimise the amount of light scattered
into particular directions without considering the light scattered
to other locations in the far-field.
This class of point-trap optimisation algorithms, or combination
algorithms, effectively combines a set of phase patterns $\Delta_m$
which describe the scattering in a particular
direction \citep{DiLeonardo2007Feb}.
In this description, we can write down a version of the Gerchberg--Saxton
algorithm which doesn't explicitly depend on the propagation method
\begin{equation}
    \phi^{j+1} = \sum_m e^{i\phi^j} \frac{V_m^j}{|V_m^j|}
\end{equation}
where
\begin{equation}
    V_m^j = \sum_{(x,y)} e^{i(\phi^j(x, y) - \Delta_m(x, y))},
\end{equation}
$(x, y)$ are the pixel coordinates and $\phi^j$ is the best guess
at the combined phase pattern.
Starting with some initial guess, such as a random superposition of
the phase patterns, the method rapidly converged.
There are extensions to the Gerchberg--Saxton algorithm for point traps
including an adaptive-adaptive algorithm and a weighted variant
\citep{DiLeonardo2007Feb}.

\section{Toolbox overview}
The Matlab toolbox is broken up into the following sub-packages:
\begin{itemize}
\item Simple patterns \verb|otslm.simple|
\item Iterative methods \verb|otslm.iter|
\item Pattern tools \verb|otslm.tools|
\item Utility functions \verb|otslm.utils|
\item Graphical user interface \verb|otslm.ui|
\end{itemize}
We also provide a directory with examples for various components of
the toolbox.
Simple patterns includes simple gratings, lenses and Hermite-Laguerre-Gauss beams.
Iterative methods include the GS 2-D and 3-D algorithms for generating a pattern
to satisfy a particular target light distribution.
Pattern tools contains functions for combining patterns and functions for visualising
the output of patterns.
Also included are a number of utility functions that we use to display the patterns
on the SLM devices with Matlab and the calibration functions.
The user interface package contains the graphical user interface components.

All the pattern generation functions take a size (for the SLM)
and output an image with double values for phase or amplitude or
complex double values for patterns with phase and amplitude.
For phase patterns, the output of these functions can take any double
value, with 0 corresponding to no phase change and 1 corresponding to
a phase change of one optical wavelength.
Before displaying or simulating these patterns, the modulo of the values followed by
scaling to a 0 to $2\pi$ range or device specific colour/voltage value is needed.
Documentation is provided on most functions in the usual Matlab way,
type \verb|help otslm| or \verb|help otslm.package.function_name|
in Matlab for information about a functions usage.
More extensive documentation is provided online and in
PDF format from the GitHub page and a ReadTheDocs page.
We would like to emphasis that with very few lines of code
it is possible to generate the figures shown in
this paper.
For example, the following is an example of the toolbox being
used to generate two kinds of beams and combine them into a
single phase pattern with a range between $-\pi$ and $\pi$.

{\footnotesize
\begin{verbatim}
% Import the library
addpath('/path/to/the/library/otslm');
import otslm.*;

% Declare the size of the device
sz = [512, 512];

% Generate a LG beam
amode = 0; rmode = 3;
lgbeam = simple.lgmode(sz, amode, rmode);
displaced_lgbeam = lgbeam + simple.linear(sz, 10, 'angle_deg', 30);
figure(1), imagesc(displaced_lgbeam);

% Generate a second beam
displaced_gaussian = simple.linear(sz, 30, 'angle_deg', -40);
figure(2), imagesc(displaced_gaussian);

% Combine the beams and convert to slm pattern
combined = tools.combine({displaced_lgbeam, displaced_gaussian});
slm_pattern = tools.finalize(combined, 'colormap', 'pmpi');
figure(3), imagesc(slm_pattern);
\end{verbatim}
} 

The patterns can be displayed on the screen using Matlab's \verb|imagesc| or similar
functions or displayed using one of the utility functions we use to apply the appropriate
colour map and display it on a region of the screen for the SLM.  These functions
are provided in the \verb|tools| directory.
The tools directory also contains the calibration functions.
The patterns can also be visualised using a near
to far field transformation (Fourier transform)
or visualised using the optical tweezers toolbox version 1 (available from
\cite{ottv1}) for non-paraxial visualisation.

A brief summary of each of the package is given in the following
section.  For more extensive details about each of the packages
we refer the reader to the toolbox documentation.

\subsection{Simple Patterns}
The initial release includes functions to generate a range of simple patterns
including:
\begin{itemize}
    \item Lenses and polynomial functions: \texttt{aspheric}, \texttt{axicon},
        \texttt{gaussian}, \texttt{parabolic},
        \texttt{spherical}, \texttt{zernike}, \texttt{cubic}, \texttt{sinc}.
    \item Periodic gratings: \texttt{sinusoid}.
    \item Common beam phase/amplitude patterns: \texttt{hgmode}, \texttt{lgmode},
        \texttt{igmode}, \texttt{bessel}.
    \item Linear ramp (\texttt{linear}) which can be used for linear gratings.
    \item Aperture functions for 2-D (\texttt{aperture}) and 3-D (\texttt{aperture3d}).
    \item A function similar to Matlab's \texttt{meshgrid} for Cartesian and
        spherical coordinates (\texttt{grid})
    \item Functions for \texttt{checkerboard}, \texttt{step} and \texttt{random} patterns.
\end{itemize}
These functions all generate Matlab arrays representing the pattern.
They can be used to describe the phase pattern of a grating, the incident
illumination (such as \texttt{gaussian}) or calculate the complex amplitude and
phase of laser pattern.
Most of these functions are only a couple of lines or use functions already
provided by Matlab.
They are provided in the toolbox with additional documentation and a
consistent interface with features specifically for SLM pattern generation.

\subsection{Iterative algorithms}
This sub-package includes a range of iterative algorithms and objective
functions that can be used with them.
The initial release includes an implementation of Gerchberg--Saxton (2-D and 3-D),
Adaptive-Adaptive, DirectSearch, and SimulatedAnnealing.
Implementations of the point-trap/combination iterative optimisation
algorithsm have been provided for Gerchberg--Saxton, Adaptive-Adaptive,
and weighted Gerchberg--Saxton.
Code is also provided to call the iterative algorithm from \cite{Bowman2017May}.
A code for optimisation in a basis of vector spherical wave functions is also provided
however this is still fairly experimental and will likely change in future releases.

The iterative methods are implemented with object orientated Matlab code.
Each iterative method is an object which inherits from a base class that provides
additional supporting methods for visualisation and fitness function evaluation.
Methods where the fitness or propagation methods are not explicitly
needed can omit these properties or choose to keep them for calculating
diagnostics/performance graphs.
The base class can easily be extended to implement other iterative algorithms.

\subsection{Tools}
The tools package contains functions for working with patterns.
This includes functions to combine patterns (\texttt{combine, mask\_regions}),
generate visualisations (\texttt{visualise}) and apply filters to patterns
(\texttt{dither, phaseblur, spatial\_filter}).

\texttt{bsc2hologram} and \texttt{hologram2bsc} functions are provided for
converting between a far-field hologram
and the Beam Shape Coefficient (BSC) beam object used in the optical
tweezers toolbox \cite{ottv1}.

\texttt{hologram2volume} and \texttt{volume2hologram} functions convert between
the 2-D hologram format and the 3-D hologram/lens format.
The 3-D hologram consists of the 2-D hologram unwrapped to a spherical
hemisphere as described in \cite{Chen2013Jan}.

\texttt{finalize} and \texttt{colormap} can be used to convert from the
0 to 1 phase format returned by most patterns to a 0 to $2\pi$ phase angle
or device specific colour-map format.
\texttt{finalize} also provides functions for combining phase and amplitude
patterns into a single phase or amplitude pattern.


    
\subsection{Utilities}
A range of utility functions are included in the toolbox.
Several functions are included for generating images of the device and
calibrating the device, some of these functions are still experimental.
A class is provided for loading/saving and interacting with device colour-maps.
Colour-maps can either be returned by a calibration function or loaded from a file.

Matlab objects are provided for representing \texttt{Showable} and \texttt{Viewable} devices
such as SLMs and cameras.
These objects inherit from two base classes, and can easily be extended to
simulate or connect to other types of devices.
Device objects for a SLM controlled by placing the pattern on a monitor
connected to the computer and a web-caemra are provided.
Additional devices are provided for simulating SLM/cameras, these are mainly used
for testing the imaging and calibration functions.

\subsection{Graphical user interface}

\begin{figure}[ht]
	\centering
	\includegraphics[width=0.8\textwidth]{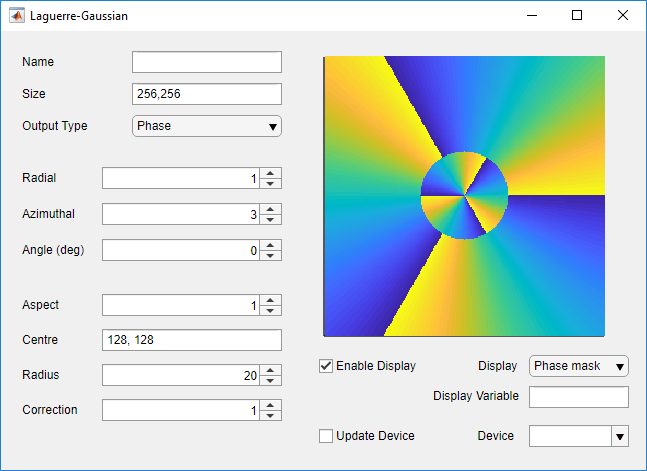}
    \caption{User interface for LG beam generation.  The interface
      provides basic controls needed to generate and display a LG
      beam.  The \texttt{Name} and \texttt{Display Variable} fields
      specify the Matlab variable names for the generated phase
      pattern and displayed image.}
    \label{fig:gui}
\end{figure}

Most of the functions in the toolbox have a corresponding
graphical user interface (GUI).
The interface is implemented in Matlab 2018a and should run on
more recent versions of Matlab.
These GUIs offer an alternative way of generating
patterns without needing to write code.
An example of a GUI for the generation of LG beams is shown in
figure \ref{fig:gui}.
The interface includes a display that can either show the current
phase pattern or the simulated far-field of the SLM. This can be useful
for quickly exploring the capabilities of a particular function.
The user interfaces are contained in the \verb|ui| sub-package.
A launcher interface has been written to provide an overview and
means to quickly launch the different components.
The user interface provides most of the same functionality as the
command line, patterns are created and added to the Matlab workspace,
allowing the interface to be used interchangeably with the other scripts.
GUI components can be launched by running them from the Matlab browser
or by calling \verb|otslm.ui.Launcher| once OTSLM has been added to the path.

\subsection{Graphics Processing Unit support}
For rapidly generating hologram patterns it is possible to use
specialised hardware, such as Graphics Processing Units (GPUs)
\citep{MartinPersson2018Sep, Bowman2014Jan}.
OTSLM currently supports two methods for utilising the GPU: generating
patterns using Matlab's \verb|gpuArray| object and directly calculating/drawing
the pattern using OpenGL.
Most functions will automatically detect if the inputs are \verb|gpuArray|
objects.
For pattern creation functions (such as \verb|otslm.simple.*| functions)
the GPU can be explicitly enabled via the \verb|useGpuArray|
optional input argument.
GPUs can be used to accelerate calculations of patterns, however, these
devices often have far less memory than the main computer.
This can create problems when, for instance, combining many large patterns
using \verb|otslm.tools.combine|.
Instead, it may be necessary to generate individual patterns
and combine them one-by-one, as is done by \verb|otslm.tools.lensesAndPrisms|.
This method can be relatively effective, using the lenses and prisms algorithm
on a GPU (tested on GTX 1060 6GB; GPUs with less memory should provide
similar performance) we were able to generate a pattern with
11 spots at 60 frames per second, faster than many of the device
we use in the lab for displaying patterns.

One of the main bottlenecks with rapidly generating simple patterns
is transferring the patterns from the computer's main memory into the
graphics memory or SLM device memory.
In some circumstances this bottleneck can be unavoidable, such as when the device
is connected via USB or to another computer on the network.
If the device is connected directly to a graphics card it is possible
use the GPU to calculate the pattern and display it directly on the device
without first copying back to the main computer memory \citep{Bowman2014Jan}.
To achieve this, OTSLM includes an interface to RedTweezers.
RedTweezers is a standalone application which draws and displays patterns
using an OpenGL enabled graphics card.
Pattern creation functions need to be described using OpenGL shader language
and compiled into shaders that are loaded onto the GPU.
Information about the pattern, such as spot locations, is then sent to the
shader and the pattern is calculated and drawn to the screen/device.
OTSLM currently includes shaders for the algorithm described in
\citep{Bowman2014Jan} as well as a shader for directly uploading images
and utility functions to help with designing a custom shader class.

\subsection{Examples}
We have tried to include a range of examples for the features included
in the toolbox.
For beam generation, we suggest the user starts with the \texttt{simple\_beams}
and \texttt{advanced\_beams} examples.
For iterative algorithms, we have provided two examples \texttt{iterative}
and \texttt{iterative\_3d} demonstrating the 2-D and 3-D algorithms currently
included.
We have also included more specific examples such as a DMD example,
different methods for generating a line trap, particle tracking using a SLM
and a demonstration to show the effect of phase blur on SLM output.
Examples are provided for calibration, imaging and setting up a screen device (window)
to control the SLM.

\section{Applications}

\begin{figure}
    \centering
    \includegraphics[width=0.9\textwidth]{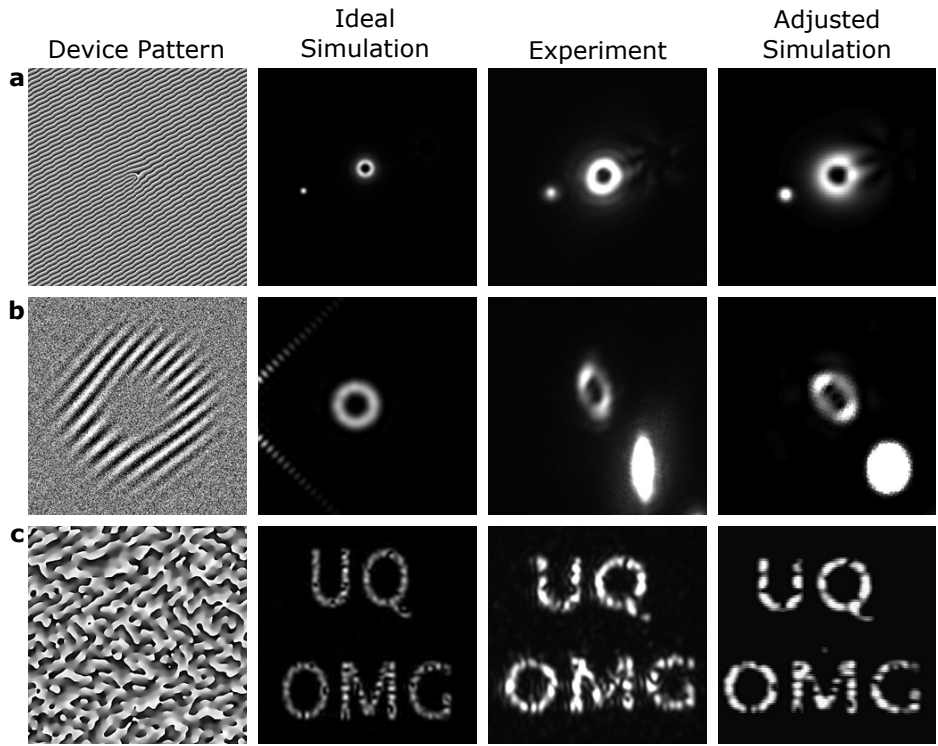}
    \caption{Examples showing three different beams generated
    using the toolbox:
    (a) a LG vortex and a Gaussian shaped spot generated using
    a single pattern on a phase-only device,
    (b) a LG vortex generated using an amplitude-only device,
    and (c) a UQ-OMG shaped light field generated using
    the Gerchberg--Saxton algorithm.
    The columns, in order, show the device pattern, the ideal
    simulated device paraxial far-field (assuming 100\%
    diffraction efficiency and no aberrations),
    what was observed in an experiment
    and the simulated device paraxial far-field after
    accounting for aberrations and imperfect diffraction efficiency.}
    \label{fig:experiment}
\end{figure}

Using the toolbox, we are able to quickly generate and simulate
a range of different optical potentials.
Figure~\ref{fig:experiment} shows a couple of examples of
different device patterns, their simulated far-field, and
the experimentally observed far-field on one of our
optical systems.
By comparing simulations of our beams to the experiment
we are able to characterise the imperfections in our
optical system.
It is then possible to either adjust our simulations to include
these imperfections (as shown in the figure), or to modify our
experiment to remove these
imperfections (either using the SLM or by modifying
other components in the system).

One of our main uses for the toolbox is generation of beams
in our optical tweezers experiments.
We have been using the toolbox for scripted control of optical
trap position and generation of optical potentials for orientating 
biological cells \citep{Lenton2019Nov}.
Being able to programmatically control where optical tweezers are placed
allows us to perform repeated measurements of fluid flows by releasing
and capturing a probe particle and measuring how far it travels.
Using this technique we can map out the velocity flow field generated
by a micro-machine or motile micro-organism.
Combining OTSLM with the optical tweezers toolbox has also allowed us to
simulate the orientation and dynamics of particles in optical traps.
We are using this method to design beams to
orientate motile micro-organisms in optical traps.

\section{Ongoing development}
We have made a public repository for our toolbox available on GitHub.
We hope that this will enable and encourage collaboration from the
community including contributions of algorithms other researchers use.
Additionally, we plan to continue to update the toolbox with new features
our group develops for optical tweezers problems.
Our main interest at the moment is optimising traps for trapping of
biological samples in arbitrary orientations.
The areas we will focus on are iterative methods and functions to
characterise and calibrate the optical systems.

Additionally, there are a couple of tasks we have not had time to
work on but we think might be beneficial.
We choose to implement the current version in Matlab for the sole
reason that our group had pre-existing code in Matlab.
It would improve accessibility and in some cases also significantly
improve runtime by switching to a non-proprietary language such as Python.
Another feature that we would consider if switching to Python is
a \texttt{Pattern} class to represent phase/amplitude patterns
returned by functions in the toolbox.
Keeping track of the pattern phase range (1 or $2\pi$) can be confusing
and sometimes lead to patterns being incorrectly scaled.
A separate class for these patterns could keep track of both the pattern
data and the other meta-data such as range, date created, functions used
to create it.

\section{Conclusion}
In this paper we have described our new beam shaping toolbox, OTSLM.
The toolbox currently includes functions to generate a range of simple
diffraction gratings and beam phase/amplitude patterns as well as
a selection of common iterative algorithms and tools for working with
patterns and devices.
Our toolbox includes a graphical user interfaces for some of the most
common tasks.
The current version of the toolbox is implemented in Matlab and
available on GitHub.
It is our hope that this toolbox will be useful to the optical
trapping community as well as more broadly in other fields where the
creation of structured light beams with SLMs is needed.
By making the toolbox open source, we hope that it will be extended
by other researchers using SLMs for light shaping.
We hope that this toolbox will be a useful resources for a wide
number of researchers in cognate fields to easily create a needed
experimental environment when sculptured light is
involved, be it in imaging or biology.

\section*{Funding}
This research was funded by the Australian Government through the Australian Research
Council's Discovery Projects funding scheme (project DP180101002).
Isaac Lenton acknowledges support from the Australian Government RTP Scholarship.

\section*{Acknowledgments}
We acknowledge the other members of the UQ optical micro-manipulation group,
particularly new members of our group who have experienced first hand the
lack of easily accessible tools for generating these patterns.
We also acknowledge Jannis Kohler and Carter Fairhall who tested early versions
of the toolbox.

\section*{CRediT authorship contribution statement}
\textbf{Isaac C. D. Lenton:} Software, Writing --- Original Draft.
\textbf{Alexander B. Stilgoe:} Writing --- Review \& Editing, Supervision.
\textbf{Timo A. Nieminen:} Funding Acquisition,
    Writing --- Review \& Editing, Supervision.
\textbf{Halina Rubinsztein-Dunlop:} Funding Acquisition, Writing --- Review \& Editing, Supervision.

\section*{References}





\bibliographystyle{elsarticle-num}
\bibliography{paper.bib}

\textbf{Pre-print of:}
\begin{quote}
    I. Lenton, et al., 
    OTSLM toolbox for Structured Light Methods,
    Computer Physics Communications,
    2020, 107199,
    \url{https://doi.org/10.1016/j.cpc.2020.107199}.
    
\copyright~2019. This manuscript version is made available under the CC-BY-NC-ND 4.0\\
license \url{http://creativecommons.org/licenses/by-nc-nd/4.0/}.
\end{quote}

\end{document}